# Mechanism of light energy transport in the avian retina

Lidia Zueva,[1,2,5] Tatiana Golubeva,[3] Elena Korneeva,[4] Mikhail Inyushin,[5]
Igor Khmelinskii,[6] and Vladimir Makarov[1]

[1] University of Puerto Rico, Rio Piedras Campus, PO Box 23343, San Juan, PR 00931-3343, USA

[2] Sechenov Institute of Evolutionary Physiology and Biochemistry, Russian Academy of Sciences, Thorez pr. 44, 194223, St-Petersburg, Russia

[3] Department of Vertebrate Zoology, Lomonosov Moscow State University, 119992, Moscow, Russia

[4] Institute of Higher Nervous Activity and Neurophysiology, Russian Academy of Sciences, Butlerova st., 5a, 117485, Moscow, Russia

[5] Universidad Central del Caribe, Bayamón, PR 00960-6032, USA

[6] Universidade do Algarve, FCT, DQB and CIQA, 8005-139, Faro, Portugal

*Corresponding Author*: Dr. Vladimir I. Makarov

[1] University of Puerto Rico, Rio Piedras Campus, PO Box 23343, San Juan, PR 00931-3343, USA

Telephone: (1)787-529-2010

E-mail: vmvimakarov@gmail.com

## *Abstract*

We consider the distribution of the intermediate filaments (IFs) in the retinal cells of the Pied flycatcher (*Ficedula hypoleuca*, Passeriformes, Aves) in the foveolar zone, which has the most specialized morphology, where the passerines have neither the rods nor the second-order neurons that service these rods, with only the single and double cones acting as photoreceptors. We report that single IFs span Müller cells (MC) lengthwise, while long cylindrical bundles of IFs (IFBs) appear inside the cone inner segment (CIS) that appearing at the OLM level. IFBs adjoin the cytoplasmatic membrane of the cone, following lengthwise at equal spacing one from the other, forming a strong skeleton of the internal cone segment, located above the OLM. IFBs follow on along the cone outer segment (COS), with single IFs separating from the IFB, touching, and sometimes passing in-between, the light-sensitive disks of the cone membrane.

Taking into account the morphology of the inverted retina in the vertebrates, we proposed a quantum mechanism of light energy transmission from the inner retinal surface the visual pigments in the photoreceptor cells. This mechanism involves electronic excitation energy transfer in donor-acceptor systems, with the donor identified with the IFs excited by the photons, and the acceptor identified with the pigments in the photoreceptor cells. The energy is mostly transferred via the contact exchange quantum mechanism. Our estimates demonstrate the energy transfer efficiencies in such systems in excess of 80%. The proposed mechanism explains the improved image contrast in fovea and foveola, and should work in daylight, while the classical mechanism that describes Müller cells as optical lightguides should predominantly operate in night vision, where some loss of resolution is traded for the ultimate sensitivity.

Our theory receives strong indirect confirmation in the morphology and function of the cones and pigment cells in the retina. In daylight conditions the lateral surface of the photosensor disks in the cones is blocked from the (scattered or oblique) light by the pigment cells. Therefore light energy can only get to the lateral surface via the intermediate filaments that absorb photons in the Müller cell endfeet and furnish the resulting excitons there. Note that the experimentally observed position of the IF extremities in the interdisk gaps is correctly reproduced in the model calculations, providing for the efficient exciton transfer. Thus, the disks are gradually consumed at their lateral surfaces, moving to the apex of the cone, while new disks are produced below. An alternative hypothesis of the non-scattered light passing through the entire cone with all of its organelles in the way and hitting the lowest disk contradicts morphological evidence, as in this case all of the other disks would have no useful function for daylight vision, never receiving much light.

# I. Introduction

We describe the application of quantum mechanics theory to the analysis of biophysical processes in a vertebrate retina. The high contrast achievable in vision requires a valid explanation, which is not provided by the classical theory describing Müller cells as optical light guides, primarily because of the small diameter of the intermediate filament bundles that transfer the light energy. However, our previously proposed quantum mechanism [1, 2] may successfully describe the transfer of the light energy to individual photoreceptor cells. This mechanism [1, 2] should be viewed as part of quantum biology, the area presently in a rapid expansion. This paper continues the development of quantum biological description of the light transmission and absorption in a retina.

Here we define some of the abbreviations to be used throughout the text:

EA – energy acceptor; ED – energy donor; El – ellipsoid; EMF - electromagnetic field; ER – endoplasmic reticulum; C – cone; CC – double cone; CIS – cone inner segment; COS – cone outer segment; IF – intermediate filament; IFB – bundle of intermediate filaments; ILM – inner limiting membrane; LD – lipid droplet; M – mitochondria; MB – Bruch's membrane; MC – Müller cell; MF – microfilament; Mt – microtubules; Mv – microvillus of MC; N – nucleus; NM – closest MV spacing; OLM – outer limiting membrane; PhR – photoreceptor; P – paraboloid; PC – pigment cell; PCp – pigment cell process; PG – pigment granule; RER - rough endoplasmic reticulum; RGC - retinal ganglion cell; Tj – tight junction; and G – Golgi apparatus.

Vertebrates have an inverted retina, with the light somehow passing through most of its thickness before getting to the photoreceptor cells located in its outer layers. Therefore, understanding the light transmission to the retinal photoreceptors requires detailed knowledge of the retinal morphology and structure of all of its layers. With all the similarity of the general outline, the retinal details may significantly differ between different groups of vertebrates, with differently structured loci distributed in a predictable pattern within the same retina. The retinal thickness may be evaluated based on the length of the Müller cells, distributed uniformly over the entire retina, with the length of these glial cells essentially equal to the retinal thickness.

Zueva et al. (2016) [1] reported a detailed morphological study of the Müller cells in an avian retina. The authors used transmission electron microscopy to demonstrate that intermediate filaments traverse the entire length of the Müller cells. They therefore suggested that the IFs are the structures responsible for the transmission of light energy by Müller cells in the avian retina [2]. These IFs begin at the MC endfoot, in contact with the vitreous body of the eye, and span through the entire MC, up to the apical ends of the MC microvilli. The authors used theoretical calculations accompanied by experiments on model physical systems [2] to demonstrate that cellular nanostructures composed of IFs may contribute to light polarization detection [3] and color separation [4] in the retina. All of this information leads to the conclusion that IFs should be responsible for light transmission in a variety of cells in different species.

Presently we report further development of the quantum theory of light interaction with the retinal structures [1-4]. The quantum mechanism of light transmission in a vertebrate retina [2-4] implies that the photons of the visible light are absorbed in the input section of the IFB adjacent to the endfoot of the Müller cell [5]. The light energy absorption causes electronic transitions in the electrically conductive IF bundle, inducing electronic excitations, excitons. The initially created excitonic wave package propagates over the conductive IFB towards its output section adjacent to the photoreceptor cells [2], with the entire energy redistribution finished in less than 3 picoseconds, with the propagation velocity close to the speed of light. Thus, the electronically excited IFB is an energy donor, with the excitons withdrawn by the energy acceptors –the photoreceptor chromophores adjacent to the output section of the IFB.

Earlier the authors reported [1] that IFs emerging from the Müller cells continue directly to the photoreceptor cells containing rhodopsin chromophores. Thus, the IFB – PhR pair affects donor-acceptor energy transfer, whereby the light energy entering through the inner retinal surface is transferred in the form of electronic excitation to the rhodopsin chromophores located in the outer layers of the retina. Presently we discuss this mechanism of the excitation energy transfer based on the detailed electron microscopy data on the retinal structure. Previously the authors reported theoretical studies of the absorption and emission spectra of the individual IFs with the incident light directed along their axis [4], finding a significant overlap between their absorption and emission spectra. These spectra contain relatively narrow spectral bands, peaking at the wavelengths dependent on their geometry and the thickness of the electrically conductive wall. Therefore, such individual IFs should be optically transparent in a narrow spectral range. However, bundle of IFs may cover the entire visible spectral range, with the resulting spectra becoming wider due to both slightly different spectral characteristics of the individual filaments and the exchange interactions between them. The authors reported were high efficiencies of the energy transfer by IFBs, achieving 90% [4] Thus, the IFs in the Müller cell – Photoreceptor assembly may explain the energy transfer in the donor-acceptor couple. Such energy transfer may be analyzed using either the contact exchange or the dipole-dipole energy transfer mechanism. In both cases, the rate constant of the energy transfer may be calculated using the golden Fermi rule [6]:

$$k_{ET} = \frac{2\pi}{\hbar}|V_{DA}|^2 \rho \tag{1}$$

here $\rho$ is the vibrational state density of the acceptor electronic excited state coupled by the $V_{DA}$ interaction to the donor electronic excited state. Here we need an estimate of the state density $\rho$ at the electronic state of the EA closest to the ED excited state. The exchange interaction is described by the expression [7]:

$$V_{DA,Exch} = e^2 \sum_{i,j \neq i} \frac{1}{|\vec{r}_i - \vec{r}_j|} \tag{2}$$

where the radius-vectors describe two electrons, one localized on the ED, another on the EA, and $e$ is the electron charge, while the dipole-dipole interaction is described by the Forster relationship [8]:

$$V_{DA,dd} = \frac{e^2}{\varepsilon \cdot r^3}\left[(r_D \cdot r_A) - \frac{3}{r^2}(r_D \cdot r)(r_A \cdot r)\right] \tag{3}$$

Here, $r$ is the distance between the centers of the respective oscillators, $r_D$ and $r_A$ distances between the centers of the oscillators and the optical electrons in ED and EA, respectively. We shall consider both mechanisms in our theoretic analysis.

The energy transfer mechanism is quite well known in photosynthetic systems [9-11]; similar principles are employed in artificial energy harvesting systems such as the fractal polymers known as *dendrimers* [12]. With chlorophyll [10] and bacteriochlorophyll [11] as chromophores, these molecules absorb photons and then act as a donor, transferring energy at nanometer distances. The chromophore molecules form a regular polygonal structure, with the acceptor located in its center. The light energy is harvested by the assembly of donor chromophores and transferred to the acceptor enzyme system, resulting in photosynthesis. A similar artificial system was built using polyphenylether dendrimers [12]. Its distributed chromophores act as EDs, supplying energy to the EA of the photosynthetic center. Thus, the IFs in the Müller cells act as energy donors, supplying the excitation energy to the rhodopsin acceptors in the photoreceptor outer segments. Note that the mechanism of the excited rhodopsin transformation is known quite well [13-15].

Presently we report detailed ultrastructural analysis of the Pied flycatcher upper retinal layers in the zone the outer limiting membrane (OLM) up and including the photoreceptor and pigment cells. We report the detailed structure of the retina, revealing the distribution of the IFs. We further present an extended theoretical model based on our earlier studies [1-4], proposing a mechanism of energy transfer from the endfeet of the Müller cells to the photoreceptor layer using on the ED – EA couples. We report numerical calculations of the ED – EA energy transfer efficiency, with the results around 80-90%. Note that the presently proposed quantum mechanism of the energy transfer via the IF energy donor to the photoreceptor energy acceptor [1-4] is a viable alternative to the conventional classical optical model [16-18], where the Müller cells are described as light-guides, similar to optical fibers. The main difficulty of such classical models is the loss of contrast due to light scattering by the internal structures of the Müller cells.

## II. Electron microscopy results

Based on our cytomorphologic studies of the Pied flycatcher (*Ficedula hypoleuca*, Passeriformes, Aves) retina, we show a diagram of a fragment of its outer part in Fig. 1. The light on this diagram comes from below, as shown by an arrow, with the part of the retina below the OLM not shown. This fragment is located in the parafoveal zone that has only single- and double-cone photoreceptors, with no rods present. It includes the layers starting from the outer limiting membrane (OLM) up the Bruch's membrane, containing the inner and the outer cone segments with the membranes carrying the photosensitive rhodopsin chromophores and the molecules of the visual excitation transfer cascade, and also the Müller cell microvilli, and the pigment cells with their branches. We were focusing on the distribution of the IFs within the cells. We measured the parameters of these cellular microstructures and the distances between them, thus obtaining a functional description of this retinal zone. Thus we followed the IFs through this zone, starting from the OLM over their entire extension to the pigment cells.

<Insert Figure 1>

At the level of the outer limiting membrane, the cone inner segment (CIS) is surrounded by the microvilli (Mv) of the two Müller cells (MC), Fig. 1. The bundles of intermediate filaments (IFB), joint the plasmatic membrane of the inner segment (CIS), along its internal surface, acting as a skeleton that provides its conical shape. Some of the IFs help organize the internal structure of the cone, concentrating glycogen of the endoplasmatic reticulum into a paraboloid shape (P), mitochondria (M) into an ellipsoid (El), and forming a lipid drop (LD) in the apical zone of the outer segment. The cone outer segment (COS), as a derivative of a cilium, starts at its base from a centriolar apparatus, with two centrioles at right angle to each other. This is the location where the light-sensitive membrane is formed, folding into a constantly growing outer segment (COS). The outer segment is surrounded by pigment cells (PC), whose processes (PCp) are filled by pigment granules (PG). Above the membranes of the outer segment and around it, parts of the cone cytoplasm, form processes extending towards the sclera, which are also containing filament bundles (IFB), with one bundle in each process. These processes extend alongside the processes of the pigment cells. Note that the outer segment is not covered by the cytoplasm completely [19]. On the open side of the membrane disks of the cone, with the plasmatic membrane absent, the outer segment is supported by thin finger-like processes (flp), whose diameter is by a factor of 5 smaller than that of the bundles.

The Müller cell structure is quite well known by now, including the existence of the intermediate filaments (IF) in its cytoplasm [5]. They extend along the entire Müller cell, from its endfoot adjacent to the inner lamina of the retina, to the ends of the microvilli above the outer limiting membrane. Initially these filaments were perceived as the mechanical skeleton, supporting the shape of the cell [20]. Later Liem (2013) [21] proposed that intermediate filaments may have additional functions, such as suppress the migration of astrocytes [22]. Recently we used the quantum confinement theory [2, 3] to justify that these IFs may also transfer the light energy, in the form of excitons, notwithstanding their very small diameter, typically close to 10 nm.

Note that according to our data obtained by light and electron microscopy, the Müller cell microvilli do not reach the light-sensitive membranes containing rhodopsin molecules directly, as they end at the distance exceeding that between the two adjacent membranes of ca. 24 nm [23]. Therefore, we assume that the light energy absorbed by the IFs and transported in the form of excitons to the zone of the outer limiting membrane (OLM) is transmitted (see Fig. 2), at least in part, to the IFs organized into cylindrical bundles (IFB), their diameter 180±20nm, which we found in the cytoplasm of the cone inner segments (CIS). These long bundles are not enveloped by a common membrane and therefore can't be classified as membrane organoids; they should rather be classified as cytoskeleton polymers. The IFBs first appear in the photoreceptor cells at the level of the outer limiting membrane (OLM), and as we are considering the foveolar retina in passeriformes that is composed of the cones only [24, 25], they extend inside the cone along the plasmatic membrane, barely touching it but providing anchorage, along the inner and the entire outer segments of the cone (see Fig. 3).

Such IFBs were not detected below the outer limiting membrane (OLM) either in the body or the synaptic pedicles of the cones [24, 28], although single IFs are found in this zone of the retina inside all of the cells along with other cytoplasmic organoids and skeletal microstructures – microtubules and microfilaments [5]. Such single IFs are distributed homogeneously within the cytoplasm and directed along the cell, repeating the bends in the cellular form [1, 2].

<Insert Figure 2>

The outer limiting membrane (OLM) has higher contrast than the surroundings in the light-microscopy preparations, being denser due to the presence of numerous microfilaments (MF) and intermediate filaments (IF). This is also the most osmeophilic part of the glial Müller cells (MC). Müller cells tightly envelop the photoreceptor cells, and due to massive presence of the intermediate filaments (IF), microfilaments (MF), and structural proteins organized into tight junctions (Tj) both between the adjacent Müller cells and between these and photoreceptors, they separate the intercellular space of the outer retinal layer from its inner layer [30]. The tight junctions (Tj) stop the free flow of the intercellular fluid between the inner and outer layer, passing only water and small molecules, including nutrients and signal molecules [25-34]. Thus, the main transport through the OLM is only possible directly through the cells [35-37].

The cones appear lighter on the electronic micrograph (Fig. 2), with cut-through microtubules (Mt), intermediate filaments (IF), Golgi apparatus (G), ribosomes, coated vesicles and other intercellular inclusions. The ribs of the cones that protrude outside are clearly visible in this thin layer; these ribs stabilize the cell location in the square pattern at the outer limiting membrane (OLM) level, limiting their rotational and translational motion. The more osmeophilic darker Müller cells send microvilli (Mv) into the intercellular space above the outer limiting membrane, carrying also one ciliary rootlet per cell in their cytoplasm.

<Insert Figure 3>

Fig. 3 shows the retinal structure on a sloped section made 2-4 μm above the level of the outer limiting membrane. Due to the slope we see both the square pattern of the double cones in the retinal plane and the dynamics of the structural changes within the cones along their height. The microvilli (Mv) form gear-like mechanical contacts with the myoids of the main and accessory cones in the basal part of the inner segment above the outer limiting membrane (OLM). The gear teeth are elongated in the ribbed part of the myoid, with the spaces between them filled by the MC microvilli (Mv). There are no microvilli at the level of the ellipsoid, the cone inner segments are slightly wider, while and their plasmatic membrane is smooth, no ribs present.

We see gray rounded formations in the same Figure inside the cones, regularly spaced along the plasmatic membrane at an approximately equal distance from each other at the side of cytoplasm. No such structures appear by the membranes separating the main and the accessory cones, as all of them are concentrated along the membranes facing the interphotoreceptor matrix, limiting the double cone in its entirety. Comparing the transverse (Fig. 3) and the longitudinal (Figure 4a) sections, we see that we are dealing with a cylindrical bundle composed of intermediate filaments. These bundles have no surrounding membrane, and are quite numerous – between nine and ten for each of the elements of the double cone.

The intermediate filament bundles (IFBs) from the two adjacent double cones form pairs in the touching points of the membranes; the pairs are separated by the membranes of the two cells, with each of the bundles in the pair located in one of the two cells in contact. These two touching bundles form the pattern of eight on the transverse section. Such configuration probably benefits the mechanical stability of the inverted retina. The presence of touching bundles provides evidence against the classic light-guide mechanism, as this mechanism would result in an efficient mixing of light following each of the bundles, thus reducing the image contrast and resolution by redistributing light to neighboring cones. As a rule, the IFs in a bundle are tightly packed into a pseudocrystalline structure, as was demonstrated for neurons [38]; because of their unique cross-bridges (side arms) and ability to align in parallel arrays, IFs are well-suited to support the extreme polar shapes of neurons [38], and expand axons for optimal electrical impulse conduction [39].

We see the same IFBs surrounding the outer segments (OS) in the upper part of the sloped section (Fig. 3), in the zone of the cross-cut outer segments. Some of IFBs pass close to the membranes, touching these by way of the cross-bridges. We know that only cone photoreceptors are present in this zone, thus we understand the asymmetry of the accompanying element distribution around the outer segment: on the side where the outer segment is surrounded by the cone cytoplasm, there are IFBs present in the latter; on the side where the outer segment is open into the intercellular space, we have the thin finger-like cytoplasmatic processes that support the cone.

<Insert Figure 4, a & b>

We see in the longitudinal section of the retina in the zone of the outer limiting membrane (Fig. 4a) the ribs of the inner segments of three cones hanging over the OLM (rib C). The left and right cones were sectioned along the ribs, which we see hanging over the belt of the outer limiting membrane (OLM), tightly strapping these cones. The cone in the centre was sectioned across the rib, and is surrounded by the apical part of the Müller cell (MC), appearing as microvilli (Mv) above the OLM. The cone cytoplasm contains microtubules (Mt), inclusions, delimited by a double membrane, and also ribosomes and polysomes. There is an intermediate filament bundle (IFB) in the left cone, oriented along the cell, starting at the cone plasmatic membrane (PM). Within the rib of the central cone we see the tight connection of the base of a similar bundle (IFB) with the plasmatic membrane. Comparing the shape, size and osmeophility of these bundles with the cross-cut rounded structures lying along the plasmatic membranes of the double cones in Fig. 3, we conclude that all of these bundles (IFBs) have similar cylindrically-shaped structures.

Notably, TEM images of Figs. 3, 4a, 4b and 5 also reveal the presence of microtubules (Mt) in the same layer of the retina. These images show the difference between the IFs and the Mts as regards their structure and thickness.

Fig. 4b shows transverse section of the retina in Pied flycatcher, showing the mechanism of the IF bundle formation, by the ribs grabbing several microvilli with subsequent appearance of the bundle in the cone internal segment directly above the outer limiting membrane (OLM). The ribs of the main element of the double cone in the zone of the outer limiting membrane surround several microvilli of the Müller cell that are close to the edge of the cluster, and elongate and extend to envelop and separate them. Thus, the bundle of filaments forms out of several microvilli of the Müller cell via formation of a closed tunnel by two cone ribs that join one another. The membrane that surrounded the bundle of micrivilli is already absent within the cone. Thus, the IFBs are formed in the retinal zone of Fig. 4b.

The interaction between IF and Mt is apparent in Fig. 5, where both IFs and Mt are clearly visible. Note that the IFB starts inside the cone from the OLM.

<Insert Figure 5>

The IF bundles, adjacent to the plasmatic membrane extend along the cone myoids, are interspaced by numerous microtubules. The myoids are filled with well-expressed rough endoplasmic reticulum (RER), framed by ribosomes. Note that all figures refer to the diagram of Fig. 1, therefore the location of the organoids and the directions are always denoted in the same way as on the diagram. The ellipsoid located above is composed of mitochondria, tightly tied together by several IFs. The ellipsoid pushes the IF bundles out of the central part of the cone, pressing them against the plasmatic membrane.

Round lipid drops are located over the ellipsoid in the main element of the double cones and also in single cones. These cellular structures have no external membrane. Lipid drops may look colorless or colored – red, orange or yellow – in fresh non-fixated total preparations. Their color is due to the carotenoids dissolved in lipids, with the optical density so high that they may work as cut-off color filters, located in the cones in the way of light before the outer segment. Our prolonged fixation made visible the filamentous material that was holding together the lipids and carotenoids, the latter were washed out during fixation and processing. Thus, the lipid drop was formed around filaments wound into a loose round ball, coming from the cytoplasm both to the lipid drops and to the lower membranes of the outer segments. The thickness of the filaments in the drops was about 20 nm; therefore each is probably composed of two twisted intermediate filaments. A cilium is visible at the upper part of the image as the level of the right cone, where the cone outer segment (COS) begins.

Fig. 6a shows the interaction of the IFs, detaching from the IFB, with the light-sensitive membrane in the photoreceptors.

<Insert Figure 6a и 6b>

A bundle of filaments (IFB), located within the same cells, extends along the cone outer segment (COS) at the left side. The distance from the bundle to the membranes of the outer segment does not exceed 80 nm. Separate disarranged intermediate filaments, detached from the bundle, are present in this gap. Small strands of filaments directly touch the disks of the outer segments, sometimes penetrating into the interdisk gap by up to 30 nm (1/70 of the disk diameter).

Dark pigment granules are visible on the level of the cone outer segment (Fig. 6b), filling the pigment cell processes (PCp). These are mobile melanin inclusions are sensitive to light intensity, and may absorb the excess EMF energy that reached the level of the pigment cells, without being absorbed by the photoreceptors cells. The pigment granules migrate upwards along the processes in low light, filling the body of the pigment cell and vacating the process, so that even weak scattered light may reach the cone outer segment. These granules move almost to the base of the outer segment at higher luminosity, reducing the scattered light and improving image contrast and resolution.

We also have the bundles (IFB) in the cone cytoplasmatic processes that surround the pigment cell processes (PCp); apparently, the energy from these bundles may be dumped onto a pigment granule via desmosome-like plackets, accelerating the photosensor system recovery. We see on the right of Fig. 6b a dark pigment granule in the pigment cell process. To the left of it, inside another process, we see an intermediate filament bundle with structures resembling half-desmosomes, adjacent to the pigment granule, and separated from it by two plasmatic membranes. The section occurred along the surface of one of these plackets (arrow) within the process to the right. We presume that these structures function to dump excess excitons onto the pigment granules. Table 1 lists typical parameters of the structures measured on the TEM images.

<Insert Table 1>

The experimental data presented here show that IFs of MCs pass through the OLM, penetrate through a special morphological system of cone ribs into the photoreceptor cell, where the IFs join each other forming the intermediate filament bundles (IFB). These bundles propagate within the entire inner segment close to the plasmatic membrane, and next within the outer segment some of the IFs get separated from the bundle. These individual IFs interact with the membrane disks, touching their surface and sometimes getting deeper into the interdisk gaps.

Next we shall use our earlier developed modeling approach [2-4], and, taking into account the experimental data reported herein, describe the mechanism of energy transfer in the inverted retina. In summary, the IFs in the MC endfeet absorb the electromagnetic radiation, generating excitons that travel along the IFBs and individual IFs; the latter enter the cones thorough openings in their membranes at the OLM level, pass through the lipid drops that function as cut-off color filters, and deliver the excitons to the photoreceptor membrane within the cone outer segment, with excess excitons dumped onto the pigment granules of the pigment cells. Physically, the energy of the electronically excited IFs is transferred to the rhodopsin molecules in the photoreceptor, with the IF – Rhodopsin acting as a donor – acceptor pair. The following section will consider this mechanism in detail.

## III. Theory of the energy exchange mechanism

Fig. 1 shows the general structure of the retina in *Ficedula hypoleuca*, shown in its scleral (outer) layer in the foveola zone, based on the morphological results presented in Figs. 2 – 6. In short, we found that continuous IFs, starting at the MC endfeet, pass through the OLM on the apical side of the MC together with the enveloping microvilli, forming bundles within the cone cells that they enter directly. The individual IFs detach from the bundle at the level of the outer cone segment, touching the membrane disks and invading the interdisk space of the photoreceptor system (Fig. 6a). Earlier we found [1-4] that IFs apparently transport light energy within the retina, with the input EMF radiation creating excited states (excitons) in the IFs, with the excitons traveling along the IFs and eventually reaching the rhodopsin in the photoreceptor membranes. Taking into account the present experimental results, we propose that the currently accepted classical mechanism [16-18], with the IFBs described as classical optical lightguides, should be complemented in daylight conditions by the quantum mechanism that infers exciton transport over continuous IFs extending through the retinal layers [1-4]. Rhodopsin is a well-known [12-15] photosensitive compound in the photoreceptor system, located in the membrane disks. Note that it is quite difficult to analyze the interaction of the IFs with the photoreceptors theoretically, due to the complexity of the system. Thus, rhodopsin may accept energy in its fundamental state, with the π-conjugated fragment in its *trans*-configuration. The initially formed singlet excited state is promptly converted into rhodopsin triplet excited state, which is further transformed into *cis*-configuration of the π-system. The photoreceptor disks hosting rhodopsin form a layered structure with the period of about 30 nm (Table 1). We show the formulas of the two rhodopsin isomers in Fig. 7 [13-15]:

<Insert Figure 7>

We shall analyze the energy transfer between the electronically excited IF and the *trans*-rhodopsin fragment shown in Fig. 7. As we already noted, the IFs may invade the interdisk space by up to 1/70 of the disk diameter (Fig. 6).

We simplified our model that includes IFs – IF bundles – separate IFs – Disks, with the scheme for modeling interactions of the IFs with the photoreceptor disk of Fig. 8. The values of the geometrical parameters shown in Fig. 8 were used in the model for the numerical analysis of the light energy transfer in the electronically excited IF – Rhodopsin pair. The *S* axis of the rhodopsin chromophore passes though the middle points of the π-double bonds (Fig. 7) parallel to the *Z* axis, while the IF is disposed along the *Z* axis, with its final stretch entering the inter-disk gap by 1/70 of the disk diameter. Table 1 lists the numerical values of the geometric parameters used in our numerical calculations.

<Insert Figure 8>

We modeled the energy transfer in the IF – Rhodopsin system using two different interaction mechanisms: (1) Short-range contact exchange mechanism and (2) Long-range electric dipole-dipole interaction mechanism.

(1) *Exchange mechanism*

As we already noted, the rate constant of the energy transfer is described by the golden Fermi rule:

$$k_{DA,Exch}(E_A) = \frac{2\pi}{\hbar}|V_{DA,Exch}|^2 \delta(E_D - E_A)$$

$$k_{DA,Exch} = \frac{2\pi}{\hbar}|V_{DA,Exch}|^2 \int_0^\infty \delta(E_D - E_A)dE_A \qquad (4)$$

$$= \frac{2\pi}{\hbar}|V_{DA,Exch}|^2 \rho_A(E_D)$$

where $V_{DA,Exch}$ is the matrix element of the exchange interaction coupling the initial system state:

$$\Psi_{DA,1} = \psi_D^{exc}\psi_A^{ground} \qquad (5)$$

with its final state:

$$\Psi_{DA,2} = \psi_D^{ground}\psi_A^{exc}, \qquad (6)$$

subscripts D and A correspond to the donor and acceptor states, superscripts "ground" and "exc" refer to ground and excited electronic states, respectively, $\rho_A(E_D)$ is the active mode state density in the electronically excited acceptor at the energy value $E_D$ of the donor excited state. Since we consider only single-electron excitations for the donor and acceptor, we limit the analysis to the two-electron exchange interaction, presented as follows:

$$\hat{V}_{DA,Exch} = \frac{e^2}{\hbar} \frac{1}{|\vec{r}_1 - \vec{r}_2|} \qquad (7)$$

that we used to analyze the D-A system of Fig. 8.

Our simplified D-A system analysis uses a series of approximations. Firstly, we neglected the exchange interaction between the IFs in the bundle, considering isolated IFs only. Next, we considered axisymmetric IFs, as shown in Fig. 9. We assumed that IFs have an electrically conductive surface layer [2-4], with the thickness $\rho$.

<Insert Figure 9>

Additionally, we considered a single *trans*-rhodopsin molecule instead of the entire photoreceptor disk, with the molecular geometry shown in Fig. 7. We assumed that both the *S* and *Z* axes intersect the *Y* axis, normal to the disk center. The distance between the *S* and *Z* axes is given by half of the interdisk gap plus half of the disk thickness, with both these values listed in Table 1. We used the angle $\gamma$ between the plane of the rhodopsin π-conjugated system and the plane containing the *S* and *Z* axes as a variable parameter. The *X* axis was directed parallel to the disk planes, while the *Y* normal to these planes (Fig. 8). Next, the location of the rhodopsin π-conjugated chromophore relative to the extremity of the IF was also a variable parameter in the calculations. Finally, we always assume an energy-resonant transition in all of our calculations. Thus, disregarding the vibrational structure of the donor, the Franck-Condon integral in the matrix element

$$V_{DA,Exch} = \langle \psi_D^{exc} \psi_A^{ground} | \hat{V}_{DA,Exch} | \psi_D^{ground} \psi_A^{exc} \rangle$$

may be determined using the relation

$$FCI = f = \langle \upsilon_A^{ground} = 0 | \upsilon_A^{exc} = 0 \rangle$$

for the fundamental 0-0 optical transition in rhodopsin.

Having evaluated the vibrational terms, we reduced our problem to the numerical analysis of the electronic matrix element

$$V_{DA,Exch}^{el} = \langle \psi_{D,el}^{exc} \psi_{A,el}^{ground} | \hat{V}_{DA,Exch} | \psi_{D,el}^{ground} \psi_{A,el}^{exc} \rangle$$

here "*el*" sub- and superscript denotes the electronic part of the respective matrix element and wave functions, the Frank-Condon integral (FCI) and the vibrational state density of the acceptor for the resonant transition $\rho_A(E_D)$. Note that we performed the numerical analysis in a simplified rhodopsin system, where the entire opsin polypeptide was substituted by a hydrogen atom. We used the Whitten-Rabinovich relationship to calculate the vibrational state density [40]:

$$\rho_A(E_D) = \frac{\left(E_v + a\sum_i \omega_i\right)^{s-1}}{(s-1)!\prod_i \omega_i} = \frac{\left(a\sum_i \omega_i\right)^{s-1}}{(s-1)!\prod_i \omega_i} \tag{8}$$

where $E_v$ is the vibrational excitation energy, $a = ½$, and $s$ is the number of vibrational degrees of freedom and $\omega_I$ are the fundamental frequencies of the normal modes.

(2) *Electric dipole-dipole energy transfer mechanism*

The rate constant of the D-A energy transfer induced by electric dipole-dipole interaction may also be presented using the golden Fermi rule:

$$k_{DA,dd}(E_A) = \frac{2\pi}{\hbar}|V_{DA,dd}|^2 \rho_A(E_D) \quad (9)$$

Reproducing the logic of the previous section, we reduce the problem to calculating the electronic matrix element

$$V_{DA,dd} = \langle \psi_{D,el}^{exc} \psi_{A,el}^{ground} | \hat{V}_{DA,dd} | \psi_{D,el}^{ground} \psi_{A,el}^{exc} \rangle$$

with

$$V_{DA,dd} = \frac{e^2}{\varepsilon \cdot r^3}\left[(r_D \cdot r_{A-}) - \frac{3}{r^2}(r_D \cdot r)(r_A \cdot r)\right] \quad (10)$$

Here, $r$ is the distance between the oscillator centers, $r_D$ and $r_A$ distance between the centers of the oscillators and the optical electrons of $D$ and $A$, respectively, according to Forster. The model in analysis, we rewrite the latter relationship as:

$$V_{DA,dd} = \frac{e^2}{\varepsilon \cdot r^3}(r_D \cdot r_{A-})[1 - 3Cos^2(\theta)] \quad (11)$$

where $\theta$ is the angle between the $z$ axis and the $r$ vector. Averaging over angle, we obtain for the interaction:

$$V_{DA,dd} = -\frac{e^2}{\pi\varepsilon \cdot r^3}(r_D \cdot r_{A-}) \quad (12)$$

We therefore obtain for the dipole-dipole interaction:

$$V_{DA,dd}^{el} = \langle \psi_{D,el}^{exc} \psi_{A,el}^{ground} | \hat{V}_{DA,dd} | \psi_{D,el}^{ground} \psi_{A,el}^{exc} \rangle \langle \upsilon_A^{ground} = 0 | \upsilon_A^{exc} = 0 \rangle$$

$$= -\frac{e^2 f}{\pi\varepsilon \cdot r^3} \langle \psi_{D,el}^{exc} \psi_{A,el}^{ground} | (r_D \cdot r_{A-}) | \psi_{D,el}^{ground} \psi_{A,el}^{exc} \rangle \quad (13)$$

Thus, we reduced the problem to the analysis of the $\langle \psi_{D,el}^{exc} \psi_{A,el}^{ground} | (r_D \cdot r_{A-}) | \psi_{D,el}^{ground} \psi_{A,el}^{exc} \rangle$ integral, and calculating $f$ and $\rho$.

## IV. Numerical analysis

Our numerical analysis used the data of Table 1. The analysis included three steps: (*a*) calculation of the interaction integral of the electronic matrix element; (*b*) calculation of the Franck-Condon integral *f*; (*c*) calculation of the state density $\rho$.

*a) Calculation of the electronic integrals*

We calculated the donor wavefunctions $\psi_{D,el}^{exc/ground}$ using the approach proposed earlier [2] and a homemade FORTRAN code [2-4]. We calculated the acceptor wavefunctions $\psi_{A,el}^{ground/exc}$ using *ab initio* approach as implemented in Gaussian-2000 package running on a Rocks Cluster system. Calculations used the coupled clusters method with 6-31G(d) basis set. We used a second homemade FORTRAN code that used the resulting wavefunctions to calculate the electronic interaction integrals for each of the two mechanisms, producing the final result in the form of the squared absolute value $\left|\left\langle \psi_{D,el}^{exc}\psi_{A,el}^{ground} \left| \hat{V}_{DA,i} \right| \psi_{D,el}^{ground}\psi_{A,el}^{exc} \right\rangle\right|^2$. These calculations used a constant diameter of the IFs equal to 10 nm (Table 1), while varying the thickness of their conductive wall. We adjusted the thickness of the conductive nanolayer for obtaining equal energies of the first electronic transition in the IF donor and in the rhodopsin acceptor. Fig. 10 shows the dependence of the calculated energy gap difference $\Delta E_{DA} = E_D^{Exc,ground} - E_A^{Exc,ground}$ on the conductive wall thickness.

<Insert Figure 10>

Fig. 10 shows that the resonance conditions occur at the conductive nanolayer thickness of about 1.9 nm. The transmission spectrum of such IFs peaks at 15,450 cm$^{-1}$ = 647 nm [4]. This thickness value was therefore used to calculate the interaction integrals. However, since the dipole-dipole interaction decays as $r^{-6}$ with the distance between the dipole centers, with *r* equal to half of the length (ca. 100 μm) of the IFs, we neglect the dipole-dipole energy transfer mechanism in our further considerations. We therefore consider the exchange mechanism only; the matrix element integral was calculated in function of the angle $\gamma$ (Fig. 8). We next averaged the integral over the angle $\gamma$ and calculated the squared matrix element *G*:

$$G = \left(\frac{e^2}{\hbar}\right)^2 \left|\left\langle \psi_{D,el}^{exc} \psi_{A,el}^{ground} \left| \frac{1}{\|r_1 - r_2\|} \right| \psi_{D,el}^{ground} \psi_{A,el}^{exc} \right\rangle\right|^2$$

in function of the location of the rhodopsin chromophore relative to the extremity of the IF. The linear size of the chromophore of the *trans*-rhodopsin molecule is about $l_{Rh}$ = 1.9 nm (Fig. 7). We illustrate the position of the *trans*-rhodopsin chromophore with respect to the IF in Fig. 7 in conjunction with Fig. 8. Note that the *S-Y* referential as shown in Fig. 7 corresponds to the plane of the π-system in the vertical position ($\gamma$ = 0, see Fig. 8). The angle $\gamma$ and the Z coordinate of the left edge of the benzene group of the *trans*-rhodopsin molecule were the variable parameters in the calculations. Fig. 11 shows the calculated results, averaged over the angle $\gamma$.

<Insert Figure 11>

Fig. 11 shows that the calculated value increases with *z*, with a tendency to saturation.

b) *Calculation of the Frank-Condon factor*

We calculated the Frank-Condon factor FCF for the 0 – 0 transition in the simplified rhodopsin molecule. The vibration wave functions $\left|\upsilon_A^{ground} = 0\right\rangle$, $\left|\upsilon_A^{exc} = 0\right\rangle$ were approximated by that of a harmonic oscillator. We used the *ab initio* method to calculate the equilibrium positions and force constants of all of the normal coordinates of the ground and excited states, as explained above, using these in the homemade FORTRAN code to calculate the FCF:

$$FCF = |f|^2 = \prod_{i=1}^{3M-6} \left|\left\langle \upsilon_{A,i}^{ground} = 0 \mid \upsilon_{A,i}^{exc} = 0 \right\rangle\right|^2$$

$$\psi_{U,\upsilon_i=0}(Q_{U,i}) = \left(\frac{\mu_{U,i}\omega_{U,i}}{\pi\hbar}\right)^{\frac{1}{4}} e^{-\frac{m\omega_{U,i}}{2\hbar}(Q_{U,i})^2} \quad (14)$$

$U = D$ or $A$

Here, $Q_{D,i} \neq Q_{A,i}$ were obtained in the *ab initio* calculation, *M* is the total number of atoms in the simplified rhodopsin molecule, as explained above. The $\mu_{U,i}$ parameters were calculated using an algorithm described earlier [41]. We obtained FCF = 0.138, using it to calculate the energy transfer rate constant in the D-A system.

*c) State density calculation*

We calculated the state density using the relationship (8), rewritten as:

$$N_v(E_D) = \frac{\left(\frac{1}{2}\sum_i \omega_{A,i}\right)^{s-1}}{(s-1)!\prod_{i=1}^{s}\omega_{A,i}} \quad (15)$$

These calculations were included in the homemade FORTRAN code mentioned in the subsection (*b*) above. The maxim number of degrees of freedom is given by $s_{max} = 3M - 6$, with $M = 62$ for the simplified rhodopsin molecule. Fig. 12 shows the calculated dependence of $N_v(E_D)$ on $s$.

<Insert Figure 12>

We used the value of 41 $(cm^{-1})^{-1}$ of the state density to calculate the energy transfer rate constant.

*d) Calculation of the energy transfer rate constant*

We calculated the rate constant was using the relationship (4). Fig. 13 shows the results that saturate at about 18 nm distance between the location of the rhodopsin chromophore with respect to the IF extremity (Figure 8). In effect, the length of the intermediate filament that contributes significantly to the exchange interaction is approximately equal to the inter-disk gap. This result provides very strong indirect evidence in favor of our theory. Indeed, the photoreceptor disks are consumed from the sides, forming the well-known cone shape. Thus, the intermediate filaments furnishing the energy of the photons to the photoreceptor disks should enter the inter-disk gap by about the gap width (15 nm), as they in fact do, producing optimum energy transfer efficiency and providing for the orderly consumption of the disks. Assuming that most of the light energy was hitting the cones directly after passing through the underlying retinal layers, the disks would be consumed in a completely different manner, namely, the bottom disk would be receiving all the energy and the other disks would have no useful function at all. What we have instead, is lateral consumption of the disks, with the light energy arriving at the cones laterally, which would be impossible unless the IFs are doing a good job, as in their absence the laterally arriving light could only be light strongly scattered within the retina, making high contrast and resolution impossible.

<Insert Figure 13>

The value of the rate constant is about $2.8\times10^9$ Hz at 18 nm. Noting that the light flux $W$ changes much slower than the reactions occurring in the vision system, we the following reaction scheme:

< Insert Scheme 1>

Scheme 1. Transformation of the energy of light, delivered by the intermediate filaments to the cone, into *cis*-rhodopsin:

1. $IF + h\nu_{exc} \rightarrow IF^*$; $W$
2. $IF^* \rightarrow IF + h\nu_{em}$; $1/\tau_{em}$
3. $IF^* \rightarrow IF$; $k_{nr}$
4. $IF^* + R_{trans} \rightarrow IF + R_{trans}^*$; $k_{DA}$
5. $R_{trans}^* \rightarrow R_{cis}$; $k_{TC}$
6. $R_{trans} + h\nu_{em} \rightarrow R_{trans}^*$; $P_{Reabs}$

here $IF^*$ is the first electronic excited state of IF, $\tau_{em}$ is its emission time, $k_{nr}$ is its radiationless relaxation rate constant, $R_{trans}$ is the *trans*-rhodopsin in its electronic ground state, $R_{trans}^*$ is its first singlet electronic excited state with a planar structure of the carbon atoms forming the conjugated π-system, $R_{cis}$ is the *cis*-rhodopsin in its ground state, $k_{DA}$ is the D-A energy transfer rate constant, $P_{Reabs}$ is the probability of reabsorption of a photon emitted by $IF^*$ and absorbed by *trans*-rhodopsin, $k_{TC}$ is the rate constant of formation of *cis*-rhodopsin ground state from excited *trans*-rhodopsin. The action of light on rhodopsin leads to the formation of *cis*-rhodopsin. The kinetic equations for $IF^*$ and $R^*_{trans}$ are given by:

$$\frac{d[IF^*]}{dt} = W - \left[\frac{1}{\tau_{em}} + k_{ne} + k_{DA}[R_{trans}]\right][IF^*] = 0$$

$$[IF^*] = \frac{W}{\left[\frac{1}{\tau_{em}} + k_{ne} + k_{DA}[R_{trans}]\right]} \quad (16)$$

$$\frac{d[R^*_{trans}]}{dt} = \frac{W\left(k_{DA}[R_{trans}] + \frac{1}{\tau_{em}}P_{Re\,abs}\right)}{\left[\frac{1}{\tau_{em}} + k_{ne} + k_{DA}[R_{trans}]\right]} - k_{TC}[R^*_{trans}] = 0$$

$$[R^*_{trans}] = \frac{W\left(k_{DA}[R_{trans}] + \frac{1}{\tau_{em}}P_{Re\,abs}\right)}{\left[\frac{1}{\tau_{em}} + k_{ne} + k_{DA}[R_{trans}]\right]k_{TC}} \quad (17)$$

and the quantum yield of *cis*-rhodopsin is given by

$$\varphi_{cis} = \frac{\left(k_{DA}[R_{trans}] + \frac{1}{\tau_{em}}P_{Re\,abs}\right)}{\left[\frac{1}{\tau_{em}} + k_{ne} + k_{DA}[R_{trans}]\right]} \quad (18)$$

As we found earlier [2,4], the transmission of the IFs may be close to 0.8-0.9, with the emission coming mostly out of the end of the IF, thus $(\tau)^{-1} \gg k_{nr}$. Taking into account the IF – rhodopsin interaction scheme of Fig. 8, we conclude that most of the emitted energy will be absorbed, therefore $P_{Reabs} \approx 1$. We therefore conclude that $1 \geq \varphi_{cis} \geq 0.8 - 0.9$, or else the quantum mechanism transfers the photon energy thought the inverted retina without significant losses. On the other hand, the quantum mechanism accounts for the high contrast of vision with the inverted retina.

## V. Discussion

### a) General Discussion

Presently we continue our work on the quantum mechanism of vision, and discuss its overall structural scheme. In particular, we examine the fine structure of the central and temporal foveas of the bifoveal retina in one-month Pied flycatcher chicks (*Ficedula hipoleuca*). It was shown earlier [35] that these specialized sharp vision zones only contain single and double cone photoreceptors, and no rods. This latter result was later confirmed by Colin (2015) for other species of passeriformes [42,44].

Tao Li et al. (2016) [44] found that the averaged retinal thickness in guinea pig is about 210±40 µm. Buttery et al. (1991) [45] reported the value of 141±17 µm for the same species. The guinea pig retinal thickness measured using histological preparations is about 128±23 µm [46]. Note that the latter value is smaller than the others. Franze et al. (2007) [16] reported guinea pig retinal thickness of 150±15 µm in freshly dissected samples. The thickness of the vascular retina in mammals measured by histological methods is 235±15 µm, comparing to 155±15 µm for avascular retina [47]. Birds typically have a larger retinal thickness as compared to mammals; indeed, the values of 350±50 µm has been reported for hawks [17]. The retinal thickness may be estimated by the length of Müller cells. The Müller cell length is 163 µm in cat and dog vascular retinas, 143 µm in humans, and 110 µm in rats. The Müller cells are shorter in avascular retinas, 79 µm in echidna and 113 µm in rabbit [18, 48-53]. All of these measurements are relevant for the explanation of the high contrast vision in an inverted retina, where the light has to pass through the scattering/absorbing retinal layers before getting to the photoreceptors. This problem has been discussed in many publications, with a review provided by Reichenbach and Bringmann (2010) [5] in their book. However, most of these publications are analyzing classic optical schemes, which do not provide a good explanation for the observations.

Earlier we proposed and discussed [2-4] a quantum mechanism of light energy transmission by the IFs of the MC, which are traverse all of the retinal layers and get in contact with the photoreceptors. We proposed that the IFs, in addition to the skeletal function, also provide for the transfer of energy within the retina. This function is based on quantum confinement operating in electrically conductive IFs; we inferred this electric conductivity on the base of the properties of macroscopic samples of proteins [2-4]. There is a common notion that the IFs are cytoskeletal and nucleoskeletal structures that provide mechanical and stress-coping resilience to the cells. The IFs also contribute to subcellular and specific biological functions, and facilitate intracellular communication [19]. However, Liem (2013) [21] discussed other functions, which IFs might also have, and our hypothetical quantum mechanism goes along with these ideas. However, there are no experimental data available on light transmission or electrical conductivity of either single IFs in the MCs or of single IFs in the eye lens. Therefore, the presently discussed quantum mechanism remains hypothetical, although with strong circumstantial evidence in the experimental morphological data and properties of nanosized conductive objects. Note that the proposed mechanism should be treated as part of quantum biology, an area in fast recent expansion. The most famous example of quantum phenomena in biology is the isotopic effect in smell [54], with the presently developed light transmission model having all possibilities to join the list.

Earlier we already discussed the similarity between the light transmission properties of the IFs and those of carbon nanotubes [4]. Indeed, two of the present authors demonstrated that vertically aligned bundles of single-wall nanotubes are highly transparent along the axis of the bundle, within a limited spectral range dependent on the nanotube diameter distribution [55]. These are exactly the same properties we expect to observe for the IFs and their bundles. Although there is no information available of the electrical and optical properties of the individual carbon nanotubes, theoretical calculations show that these must be very similar to those of nanotube bundles. Thus there is still the need to explore the properties of both the individual IFs and the individual carbon nanotubes.

b) *Morphology of the Ficedula Hipoleuca retina*

The presently explored retinal zone in the Pied flycatcher (*Ficedula hypoleuca*, Aves, passeriformes) has a unique structure, as it contains the cones only, with no rods present. We found that the IFs traverse all of the retinal layers (Figs. 1 – 6), starting from the ILM and directly interacting with the photoreceptor disks. Taking into account these morphological results, we proposed that the IFs have an additional function as guides of the light energy. This we have a donor-acceptor system, where the IFs are absorb light, passing into their excited electronic state, and transfer this excitation (exciton) to the photoreceptor (acceptor) by the energy exchange mechanism. We deduced such properties of the IFs in the inverted retina based on our previous work on the quantum mechanism and indirect experimental evidence that uses model systems [2-4]. As the same physical equations that are valid for the metallic nanowires are also controlling the IFs, we also expect a high similarity between the properties of the former and the latter. Although these properties were not demonstrated experimentally in the IFs, there exist various additional functions of the IFs [27, 51, 57-60], apart from the skeletal functions [53]. Indeed, a recent work elucidated that genetic defects result in structural alterations in the IFs that affect their involvement both in signaling and in controlling gene regulatory networks [27]. The paper highlights the basic structural and functional properties of the IFs and derives a concept of how mutations may affect cellular architecture and thereby tissue construction and physiology.

Our TEM images show IF structures unknown earlier in the cones of avian eyes. The issue of the light transmission by the MCs in the inverted has been discussed for over 10 years [5]; as we already noted, we used the quantum confinement theory [2-4] to infer that the non-polar intermediate filaments, ca. 10 nm in diameter, function as the light energy conductors between the ILM and the photoreceptor system.

The presently discussed morphology of the central and temporal foveas in the Pied flycatcher and its comparison to the studies allowed to us to understand the role of the quantum mechanism in the vision. The classical optical mechanism has been extensively discussed earlier [58]. The authors describe light focusing by the cellular organelles to explain the increased image contrast, as apparently confirmed by microwave model studies [58]. However, the optical mechanism can't produce focusing of light scattered within the retina, and can't explain the high-contrast vision. We therefore believe that the quantum mechanism provides for the high-contrast vision in daylight, when the pigment cells are absorbing the light scattered within the retina, while at night the high contrast and resolution are traded for the ultimate sensitivity provided by the classical optical mechanism, making use of all of the available light.

*c) Intermediate filament structure*

Much attention was given to the IF structure and functions [53-56]. However, none of these studies address the optical properties of the IFs in the cells of transparent tissues. Still, we should use this information when discussing the properties of the IFs located in MCs.

The role of the IFs in the cytoskeleton and nuclear boundary is well known, along with their role in the functional organization of the cellular structural elements. Morphologically similar while biochemically different IF networks with specific nanomechanical functions appear in different cell types. As already noted, the new functions we attribute to the IFs are not excluding any of the earlier ideas. The structure of the vimentin central α-helical domain and its implications for the intermediate filament assembly, and the IF dimer atomic structure were reported in detail earlier [27, 54, 57]. As we already noted, the IFs may assemble into dimers and more complex structures, therefore the presently observed IF bundles do not contradict the earlier reported results [27, 60, 62].

  *d) Physical aspects of the IF conductive nanolayer/*

Note that no direct information on the electric conductivity of the IFs is available at this time. Nevertheless, many polypeptides do exhibit very high electric conductivity [63], as also reported in a review [64]. Therefore, we presume that the specialized IFs are in fact electrically conductive, thus our theoretical models should provide a valid description of light transmission by the IFs [2, 3]. The experimentally observed electrical conductivity of such polypeptides was explained by the electrical conductivity of a common conjugated π-system, distributed over the entire biopolymer molecule and including carbon and nitrogen atoms. The electronic state structure of such biopolymers depends on the interactions of the conjugated π-systems of its distinct parts, such as α-spirals, and should have qualitative similarities with that of the multilayer carbon nanotubes, where the energy gap between the bonding and antibonding orbitals depends on the carbon nanotube diameter and number of layers [61-63]. We therefore assume that the IFs in the Müller cells may have the same type of molecular structure involving extended conjugates π-systems interacting with each other and providing the high electric conductivity of the IFs. The $\pi$-$\pi^*$ transitions in such systems are typically located in the red and middle-IR spectral regions, while the $n$-$\pi^*$ transitions are typically located in the near-IR, visible, and ultraviolet spectral regions [65-67]. These latter transitions are similar to the transitions generated by the quantum confinement, as discussed earlier [2-4] and presently. Therefore, the known structure of the IFs in Müller cells is compatible with the idea of electrically conductive rod-shaped biopolymer assemblies used in our description of light transmission.

  *e) Mechanism of the energy exchange process*

Here we proposed the energy exchange mechanism between an electronically excited IF (energy donor) and a photoreceptor, containing the *trans*-rhodopsin chromophore (energy acceptor). Using numerical analysis of the energy transfer mechanism, we deduced that the contact exchange interaction of the rhodopsin molecule is much smaller than that of the IF, and the rhodopsin molecule is located in the output zone of the IF, the dipole-dipole energy exchange mechanism in unimportant, as the perturbation induced by this mechanism is proportional to $1/r^3$, where $r$ is the distance between the centers of the donor and the acceptor, which results in very small values, given the length of the IF. This leaves us with the exchange interaction as the main mechanism providing for the excitation transfer between the IF and the photoreceptor, and with the need to account for the possibility of the exchange interaction operating at distances as large as 12.5 nm, with the donor and acceptor separated by the disk membrane. The presence of the membrane was disregarded in our analysis, and here we shall describe its effects on the exchange interaction. The exchange interaction is a purely quantum phenomenon, occurring in systems with 2 or more electrons [68]; we were considering only two electrons, one in the donor and another in the acceptor. Adding the disk membrane to the picture, we must add the third electron, present in the membrane. Thus, the strength of the interaction between the donor and acceptor, mediated by the membrane, will depend on the energy gaps in the three components. Provided the three energy gaps are comparable, we may expect stronger exchange interaction due to reduced distances between the components.

*f) Light energy transfer dynamics*

We considered the mechanism of light energy transfer by the IFs, and used the transfer efficiency coefficient of 0.8-0.9, as predicted by our models [4]. The light energy is used for formation of *cis*-rhodopsin, as shown by the reactions 1 – 6, listed in Scheme 1. This scheme provides high quantum yields of *cis*-rhodopsin; however, we omitted the radiationless relaxation for the $R^*_{trans}$, species, which might significantly affect the *cis*-isomer yield. Smitienko et al. (2014) [14] found using femto-photochemistry research technology that quantum yield of the reversible photoinduced *cis-trans* rhodopsin transformation is close to unity. Therefore, the presently proposed mechanism predicts correct values for the rhodopsin isomerization quantum yield, and may be used in future theoretical developments.

*g) General consideration on quantum phenomena in biologic systems*

Presently we discussed the quantum mechanism of light energy transfer in an inverted retina. However, presently the researchers come to a wider understanding of the role of quantum phenomena processes in biological systems [69]. The key role of quantum mechanics in the origin and operation of biological organisms is coming to light, beyond the trivial notion of defining the shapes and sizes of biological molecules and their chemical affinities. These range from Schrödinger's suggestion that quantum fluctuations produce mutations, to Hameroff and Penrose's conjecture that quantum coherence in microtubules is linked to consciousness. Davies, (2004) [69] reviewed some of these ideas, discussing the serious problem of decoherence. He advanced some further conjectures on the quantum information processing in biologic systems, suggesting some experiments. The presently proposed mechanism for the light energy transmission by the IFs is directly related to the issues discussed by Davies. Here we did not address the coherence of the originally prepared excited wave-package and its dephasing dynamics, as we discussed these issues earlier [2]. We reported the dephasing does not affect the emission quantum yield, being therefore of no relevance [2]. On the other hand, we did all further calculations in quasistationary conditions, analyzing time-independent probability of the energy transfer by the IFs [3,4] and in the donor-acceptor system (current study), including the IF – photoreceptor chromophore donor-acceptor couple. Thus, the presently proposed mechanism clearly belongs to the field of quantum biology, providing a viable explanation for high-contrast vision in daylight.

## VI. Experimental Methods

Histological tissues were collected in the Moscow region (Russia). Wild Pied flycatcher nestlings (*Ficedula hypoleuca*, Passeriformes, Aves) were taken from the nests at the second part of the nesting period and raised in artificial conditions. At the age of 10, 14 or 27 days after hatching, the eyeballs were incised under dim red illumination and placed in the fixative: 2.5% glutaraldehyde, 4% paraformaldehyde in 90 mM sodium cacodylate buffer with 0.02 mM $CaCl_2$, pH 7.4, and stored for a month at +4°C. The preparations were sectioned into nasal, central, and temporal regions, washed separately with 90 mM sodium cacodylate buffer, postfixed in 1% osmium tetroxide ($OsO_4$) with 1.5% $K_4[Fe(CN)_6]$ for 30 minutes in the same buffer, washed, incubated in 1% $OsO_4$ for 30 min, washed in distilled water, and then incubated in a 2% aqueous solution of uranyl acetate $[UO_2(CH_3COO)_2·2H_2O]$ for 1h and washed. After dehydration through a graded series of acetones, the tissues were oriented relative to the position of the pecten and embedded in EMbed 812 EMS epoxy resin (all of the chemicals were ordered from Electron Microscopy Sciences, Hatfield, PA, USA). Ultrathin sections 60 nm thick were prepared using a LKB Ultratome (LKB-Produkter, Bromma, Sweden) and examined with JEM 100B and JEM 1011 electron microscopes (JEOL Ltd., Tokyo, Japan).

# VII. Conclusions

In the present study we provided a detailed analysis of the morphology of the scleral retinal layer of the Pied flycatcher, starting from the outer limiting membrane (OLM) and including the Müller cell microvilli, cone outer and inner segments, and also the pigment cells up to the Bruch's membrane, with a special attention to the distribution of the intermediate filaments in the cells. We report that the IFs passing out of the MCs get combined into bundles (IFBs) that continue inside the cones. Finally, single IFs separated from the IFBs, entering the gaps between the photoreceptor disks. Taking into account the morphology of the inverted retina, we proposed a quantum mechanism of light energy transfer from the internal retinal surface (ILM) to the photoreceptor system. This mechanism used energy transfer between the electronically excited IF (donor) to the ground-state chromophore in the *trans*-rhodopsin molecule (acceptor). This energy transfer should operate via the contact exchange quantum mechanism. The calculations estimate the energy transfer efficiency in excess of 80-90%. The proposed mechanism quantum mechanism operates in daylight, providing high contrast and resolution, while the classical optical mechanism should operate at low light levels, providing highest possible sensitivity at lower contrast, due to scattered light. This scattered light is cut off by the pigment cells in the daylight, reducing the total light flux but improving the contrast.

Our theory receives strong indirect confirmation in the morphology and function of the cones and pigment cells in the retina. In daylight conditions the lateral surface of the photosensor disks in the cones is blocked from the (scattered or oblique) light by the pigment cells [25]. Therefore light energy can only get to the lateral surface via the intermediate filaments that absorb photons in the Müller cell endfeet and furnish the resulting excitons there. Note that the experimentally observed position of the IF extremities in the interdisk gaps is correctly reproduced in the model calculations, providing for the efficient exciton transfer. Thus, the disks are gradually consumed at their lateral surfaces, moving to the apex of the cone, while new disks are produced below. An alternative hypothesis of the non-scattered light passing through the entire cone with all of its organelles in the way and hitting the lowest disk contradicts morphological evidence, as in this case all of the other disks would have no useful function for daylight vision, never receiving much light.

Note that some similar engineering solutions were already proposed, with a retinal implant based on a Si nanowire matrix. This system was developed at University of California San Diego and La Jolla-based startup Nanovision Biosciences Inc. [70]. The researchers demonstrated neuronal response to light in a rat retina interfaced to a device prototype in vitro. We believe that the current study may produce useful insights that could help rethink the engineering solutions.

**Acknowledgements:** the authors are grateful for the financial support of the RSF grant # 16-14-10159 for L.Z., PR NASA EPSCoR grant # NNX13AB22A for V.M., NIH grant # SC2GM111149 for M.I. and RSF grant # 17-06-00404 for E.K.; T.G. thanks G.N. Davidovich and A.G. Bogdanov (Electron microscopy laboratory, Faculty of Biology, Lomonosov Moscow State University) for technical assistance.

## Author Information

1. University of Puerto Rico, Rio Piedras Campus, PO Box 23343, San Juan, PR 00931-3343, USA

   Lidia Zueva, Vladimir Makarov

2. Sechenov Institute of Evolutionary Physiology and Biochemistry, Russian Academy of Sciences, Thoreza pr. 44, 194223, St-Petersburg, Russia

   Lidia Zueva

3. Department of Vertebrate Zoology, Lomonosov Moscow State University, 119992, Moscow, Russia

   Tatiana Golubeva

4. Institute of Higher Nervous Activity and Neurophysiology, Russian Academy of Sciences, Butlerova st., 5a, 117485, Moscow, Russia

   Elena Korneeva

5. Universidad Central del Caribe, Bayamón, PR 00960-6032, USA

   Lidia Zueva, Michail Inyushin

6. Universidade do Algarve, FCT, DQB and CIQA, 8005-139, Faro, Portugal
   Igor Khmelinskii

## Contributions

L.Z., T.G., and E.K. designed and performed experiments and analyzed data; V.M., I.K. designed, performed and analyzed theoretical model, developed application of the theoretical model to the experimental data analysis; I.K., V.M., L.Z., T.G., M.I., and E.K. contributed to scientific discussions. V.M., L.Z., T.G., and I.K. wrote the manuscript with input from all coauthors.

## Corresponding author

Correspondence to Dr. Vladimir Makarov.
Phone: (1)787-529-2010
E-mail: vmvimakarov@gmail.com

## Figure captions

Figure 1. Diagram of the foveolar zone of the scleral (outer) retinal layer in Pied flycatcher (*Ficedula hypoleuca*). The cone inner segment (CIS) is surrounded by microvilli (Mv) of the two Müller cells (MC). The cone outer segment (COS) with its photosensitive membranes is in contact with the pigment cells (PC), whose processes (PCp) are filled with pigment granules (PG); position of PG is shown generically and does not refer to any specific value of light intensity. The vertical axis of the diagram coincides with the *z* axis of the space-fixed Cartesian referential. The light comes from below, from the inner limiting membrane (ILM, not shown) towards the pigment epithelium, as shown by the arrow in the bottom.

Figure 2. Transverse retinal section of Pied flycatcher on the level of the outer limiting membrane (OLM). C – cone, CC – double cone, MC – Müller cell, Mv – Müller cell microvilli, MT– microtubules , Tj – tight junction.

Figure 3. Oblique transverse retina section of Pied flycatcher 2 - 4 μm above the OLM. OLM – outer limiting membrane, CC – double cone, Mv – Müller cell microvilli, M – mitochondrion. Tj – tight junction, ribs, cilia.

Figure 4. (a) Longitudinal section of the Pied flycatcher retina with the OLM. (b) Transversal section of the Pied flycatcher retina with the OLM. Cone ribs envelop the MC microvilli, creating a duct for the IFs. MC microvilli enter the cone cytoplasm. IF bundles form in the cone cytoplasm.

Figure 5. Longitudinal section of Pied flycatcher retina in the apical part of inner segment at the ellipsoid level.

Figure 6. Cone outer segment (OS) composed of lipid membrane disks with molecules of rhodopsin, the photosensitive protein. Longitudinal section: (*a*) – general view; (*b*) – left side; (*c*) – right side.

Figure 7. Chemical formulas of *cis*- and *trans*-rhodopsin isomers. The numbered carbon atoms show the place where cis-trans isomerization takes place. The cartesian referential specifies the spatial orientation of the *trans*-rhodopsin molecule in Fig. 8.

Figure 8. Model geometry for the exchange interaction calculations in the intermediate filament – rhodopsin energy transfer D-A couple. See Fig. 7 for the spatial orientation of the *trans*-rhodopsin molecule. In model calculations, the left edge of the benzene ring swept increasing $Z$ coordinate values starting from $z_{IF}$.

Figure 9. Hypothetical geomentry of the intermediate filament used in the model calculations.

Figure 10. Dependence of the calculated value of the energy gap $\Delta E_{DA} = E_D^{Exc,ground} - E_A^{Exc,ground}$ on the conductive nanolayer thickness. The intermediate filament length is 100 μm, $\underline{R}$ = 5 nm.

Figure 11. Dependence of the interaction matrix element value (*G*) on the rhodopsin chromophore location (*Z*-coordinate). The results are averaged over the angle *γ*.

Figure 12. State density plot in the rhodopsin molecule, with the opsin moiety substituted by an H atom.

Figure 13. Dependence of the energy transfer rate constant on the rhodopsin chromophore location (*Z*-coordinate). The results are averaged over the angle *γ*.

Table 1. Typical dimensions of the elements shown on the scheme of Fig. 1.

| Object/parameter measured | Diameter/thickness, nm |
|---|---|
| Diameter of Intermediate Filament, IF | 10.0±1.7 |
| Diameter of a bundle of intermediate filaments (IFB) | 180±20 |
| Diameter of Microvilli (Mv) in MC | 120±20 |
| Nearest MV distance (NM) | 10 – 20 |
| Diameter of microtubules (MB) | 26.0±1.2 |
| Diameter of pigment cell process (PCP) | 95 – 300 |
| Diameter of pigment granule (PG) | 200 |
| Disk thickness (DT) | 32 |
| Diameter of side arms (cross-bridges) | 4.8±1.8 [38] |
| Length of side arms (cross-bridges) | 34±10 [38] |
| Müller cell (MC) length | ≈ 300 μm (equal to thickness of the Pied flycatcher retina in the central parafovea) |
| Average diameter of cone outer segment (COS) | 0.92±0.04 μm |

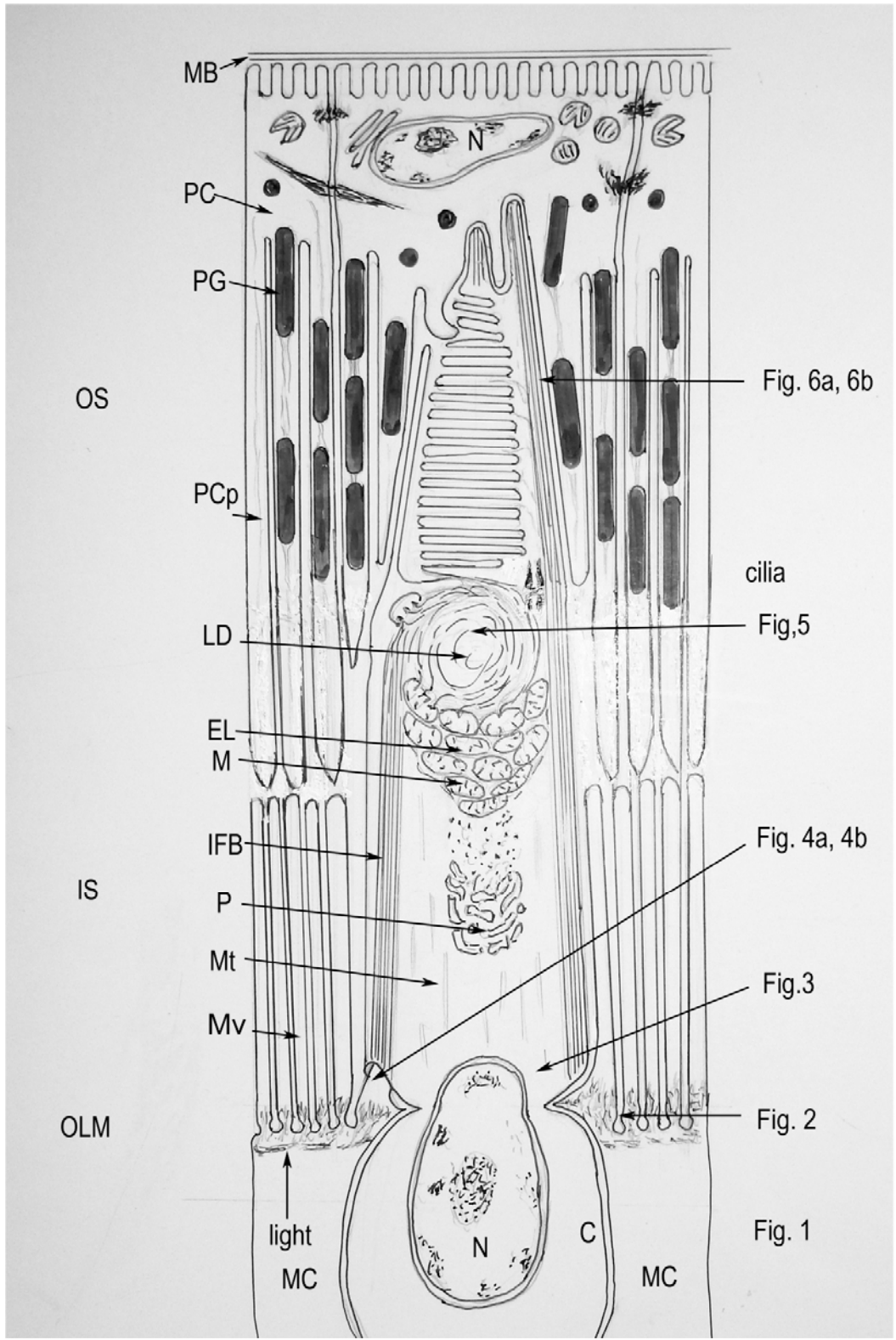

Figure 1. Makarov, et al.

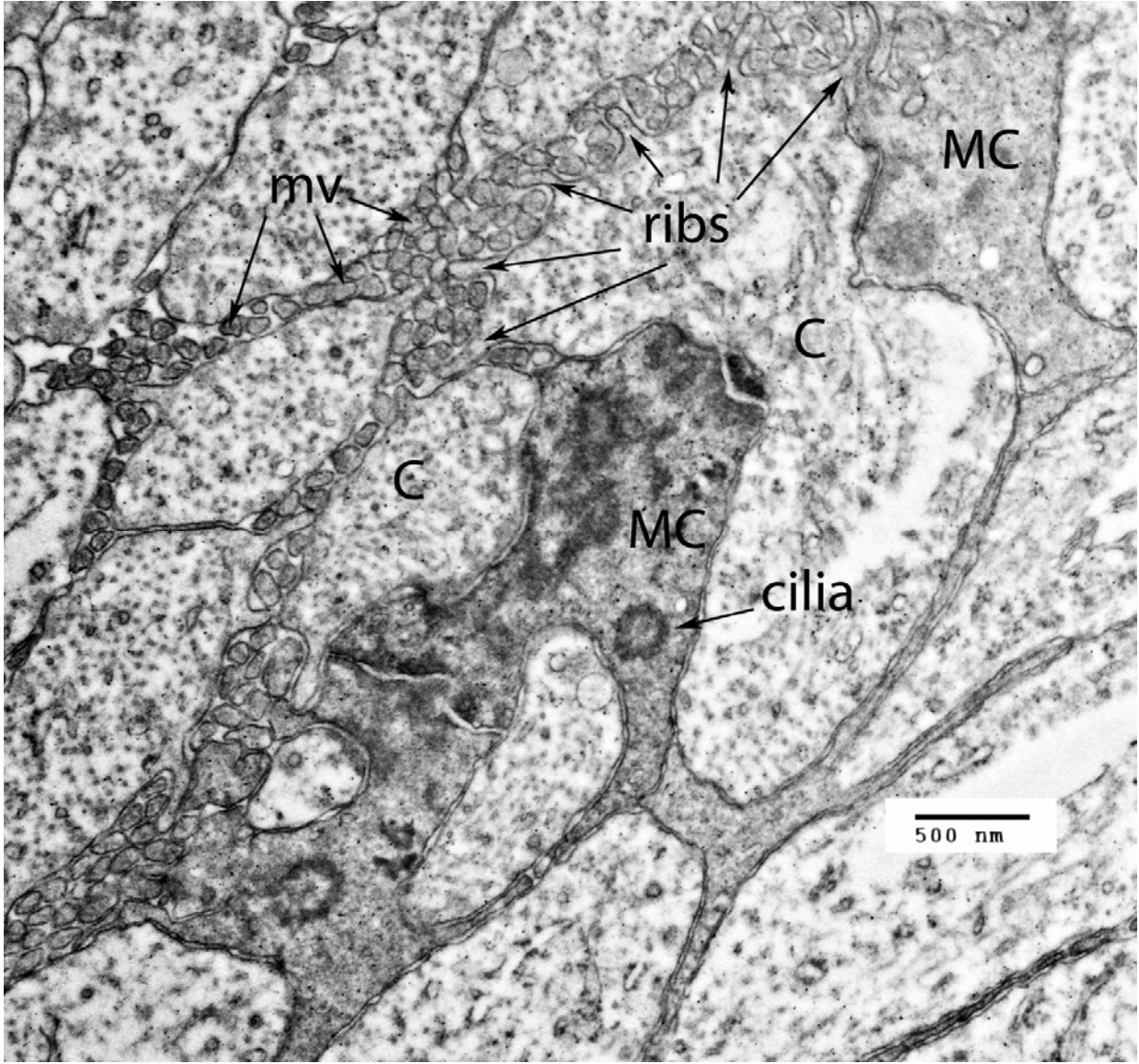

Figure 2. Makarov, et al.

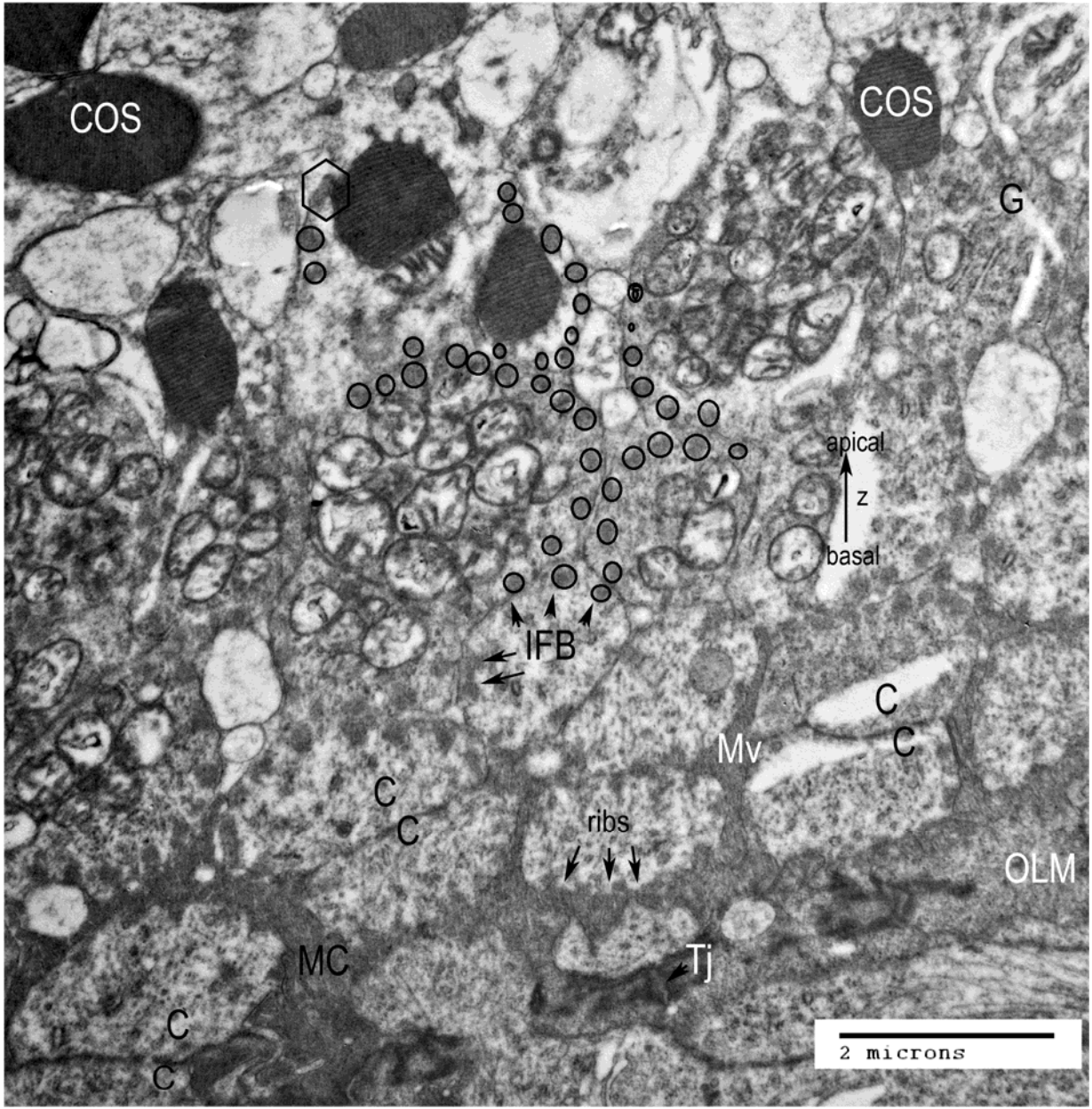

Figure 3. Makarov, et al.

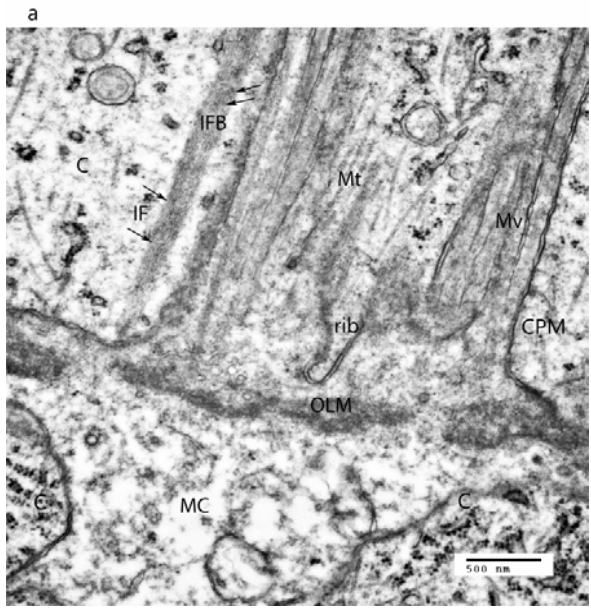 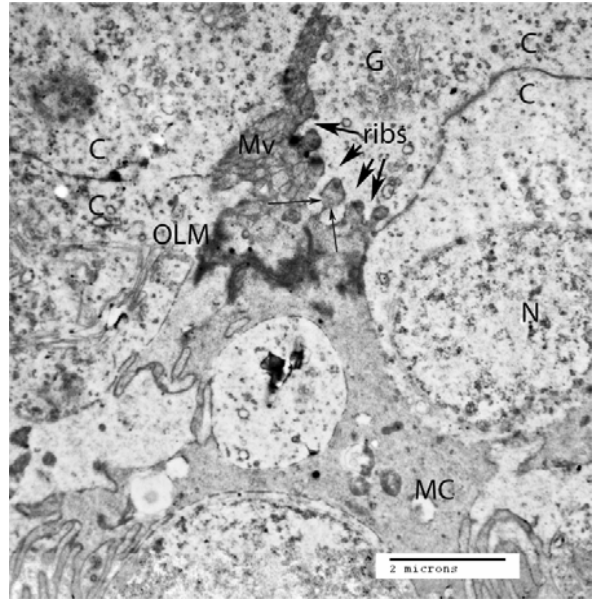

Figure 4a and 4b. Makarov, et al.

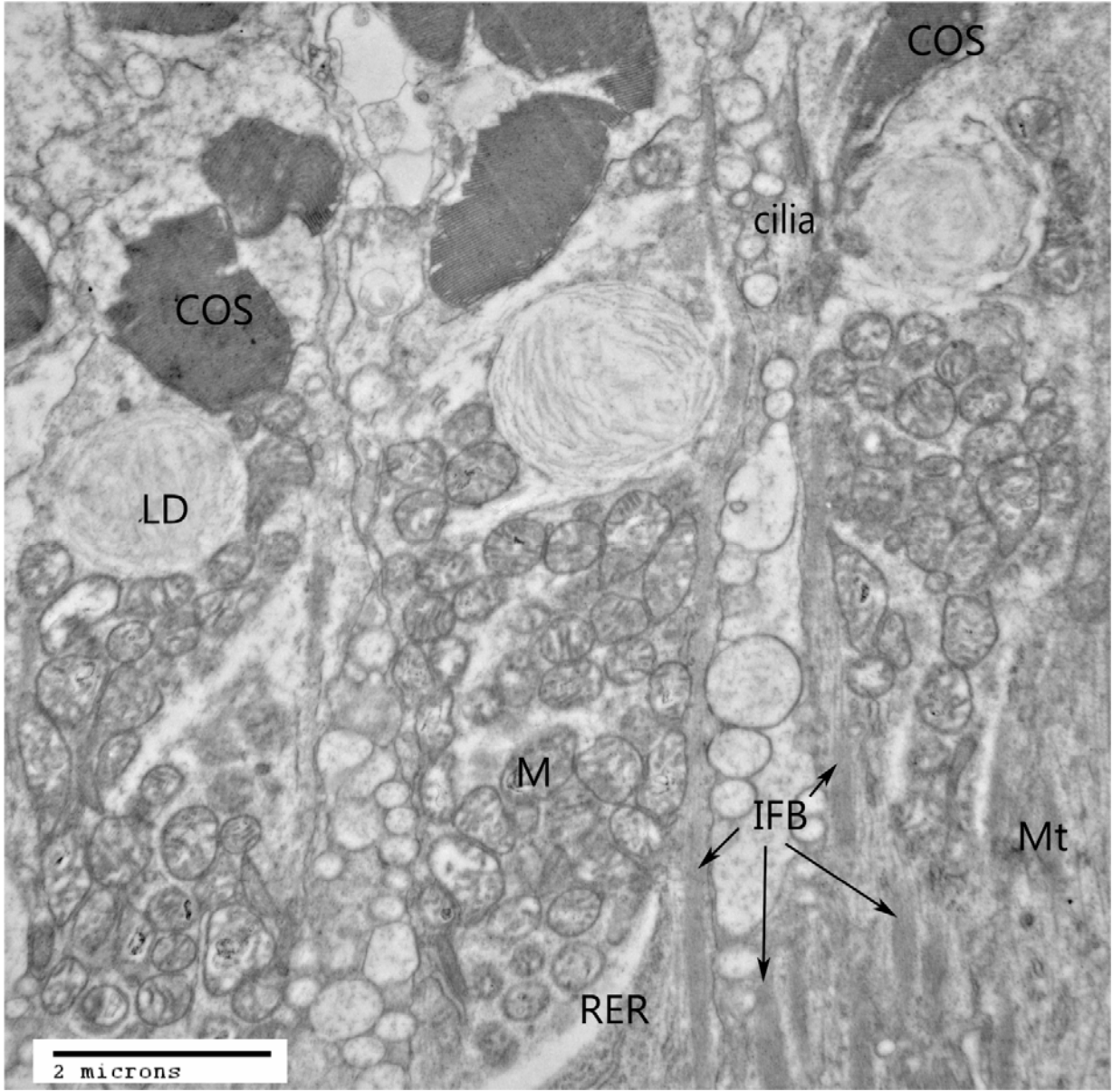

Figure 5. Makarov, et al.

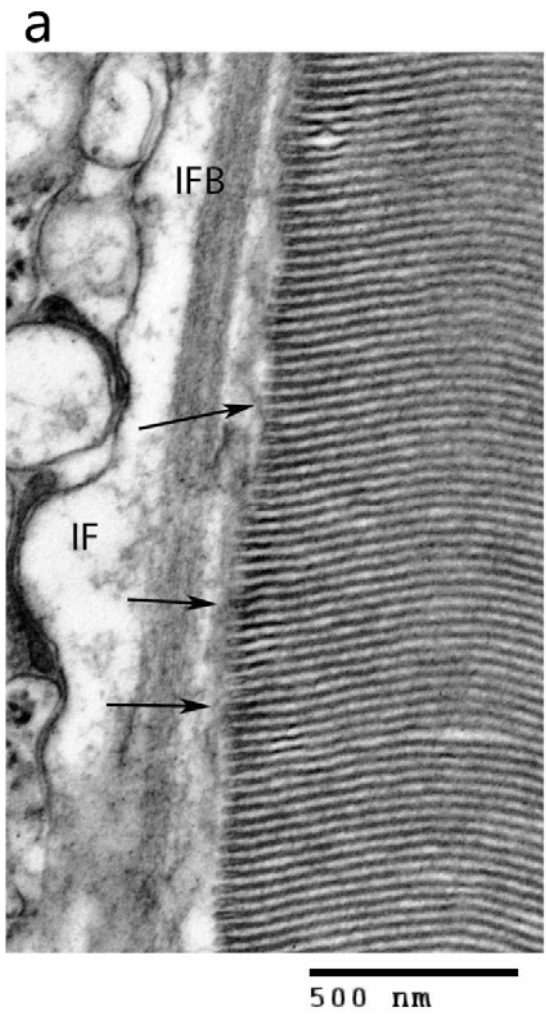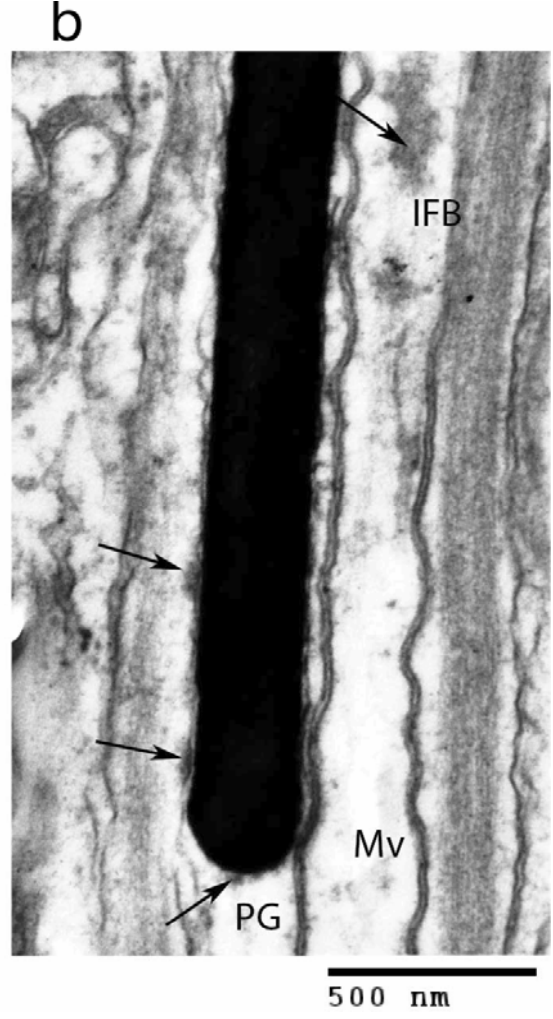

Figures 6a and 6b. Makarov, et al.

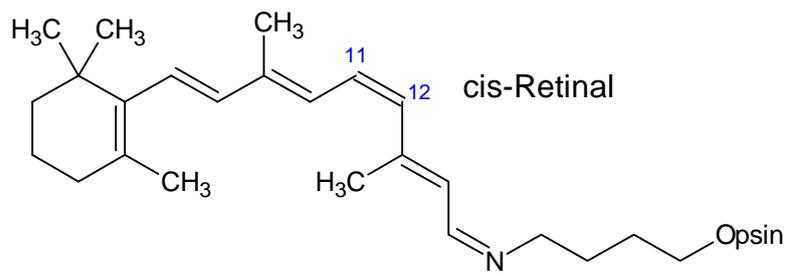

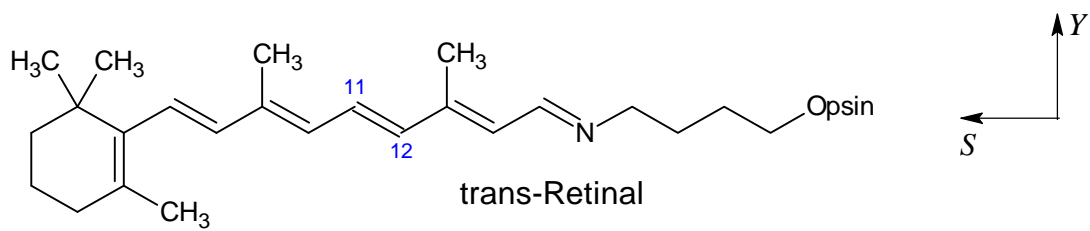

Figure 7. Makarov, et al.

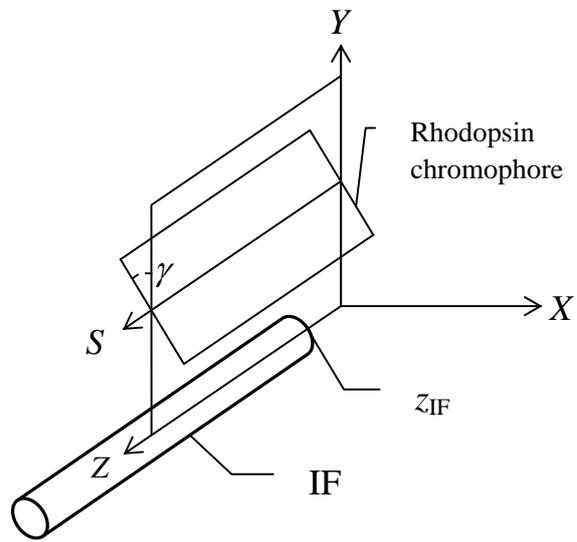

Figure 8. Makarov, et al.

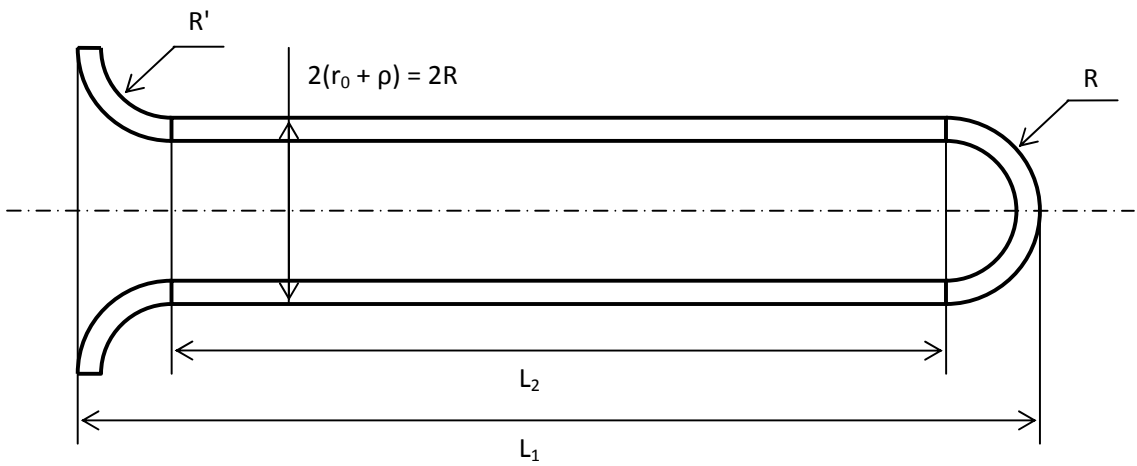

Figure 9

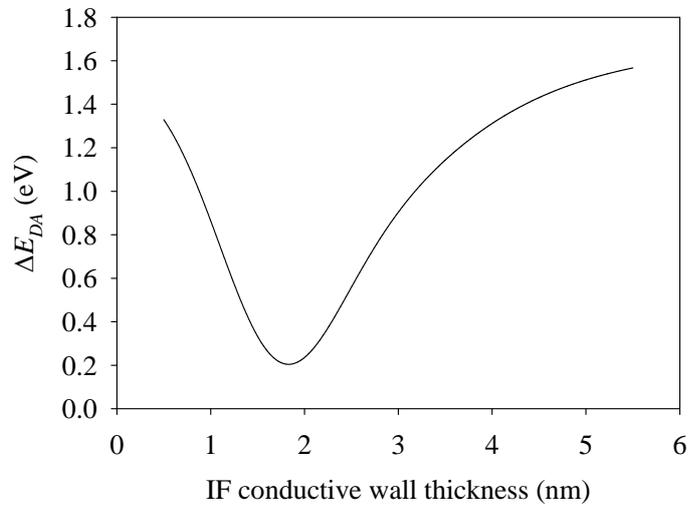

Figure 10. Makarov, et al.

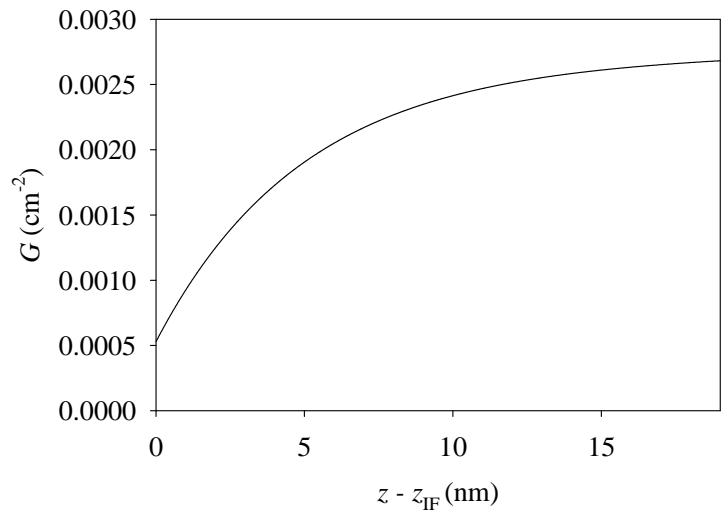

Figure 11. Makarov et al.

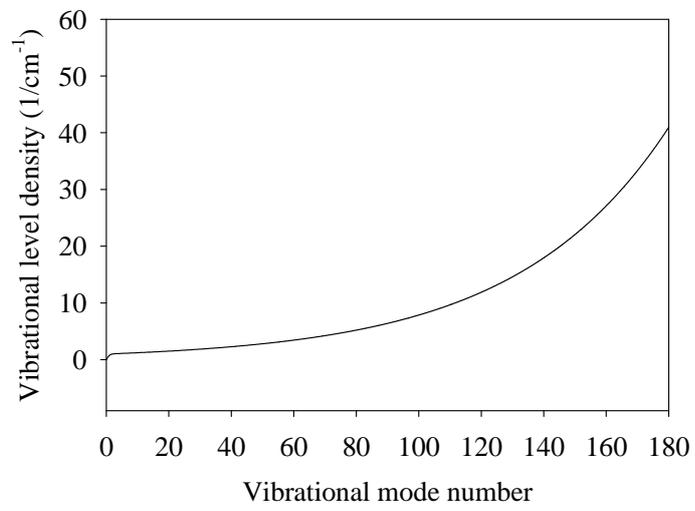

Figure 12. Makarov, et al.

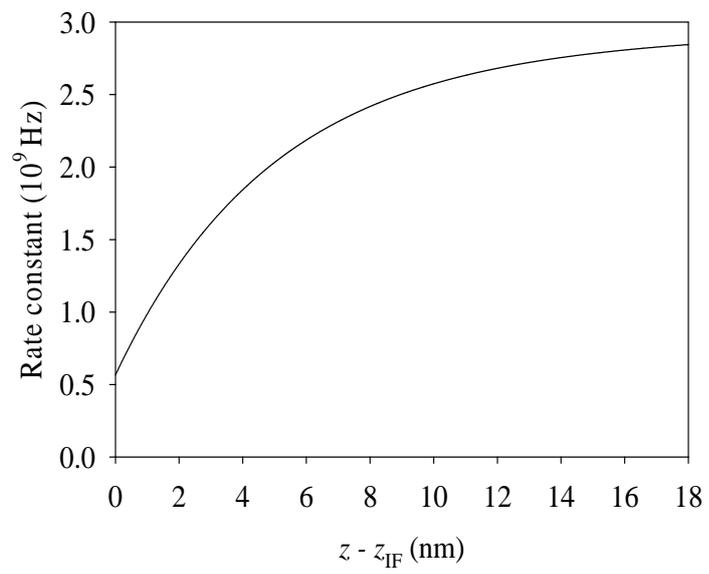

Figure 13. Makarov, et al.